\documentclass[useAMS,usenatbib]{mn2e}
\usepackage{graphicx}
\usepackage{lscape}
\usepackage{ulem}
\usepackage{times}

\def\kms{km~s$^{-1}$}
\def\cm{cm$^{-2}$}
\def\lya{Ly$\alpha$}
\def\nhi{$N$(H{\sc \,i})}

\def\ebmv{$E(B$$-$$V)$}

\newcommand{\bspsmall}{\vspace{0.5cm}\small\noindent This paper has been typeset
from a \TeX/\LaTeX\ file prepared by the author.\normalsize}

\title[DIBs in $z<0.6$ Ca{\sc \,ii} Absorbers]{Diffuse
  Interstellar Bands in $\bmath{z < 0.6}$ Ca{\sc \,ii}
  Absorbers\thanks{Based on observations made with ESO Telescopes at
    the Paranal Observatories under programme ID 077.C-0737(A)}}

\author[Ellison et al.] {Sara L. Ellison$^1$, Brian A. York$^1$, Michael T.
Murphy$^{2,3}$, Berkeley J. Zych$^2$, 
\newauthor Arfon M. Smith$^4$ and Peter J. Sarre$^4$\\
$^1$Department of Physics and Astronomy, University of Victoria, Victoria, B.C., V8P 1A1, Canada\\
$^2$Institute of Astronomy, University of Cambridge, Madingley Rd., Cambridge, CB3 0HA, UK\\
$^3$Centre for Astrophysics \& Supercomputing, Swinburne University of
Technology, Hawthorn, Victoria 3122, Australia\\
$^4$School of Chemistry, The University of Nottingham, University Park, Nottingham, NG7 2RD, UK}

\begin{document}

\pubyear{2007}

\maketitle

\label{firstpage}

\begin{abstract}
  The diffuse interstellar bands (DIBs) probably arise from complex
  organic molecules whose strength in local galaxies correlates with
  neutral hydrogen column density, \nhi, and dust reddening, \ebmv.
  Since Ca{\sc \,ii} absorbers in quasar (QSO) spectra are posited to have high
  \nhi\ and significant \ebmv, they represent promising sites for the
  detection of DIBs at cosmological distances.  Here we present the
  results from the first search for DIBs in 9 Ca{\sc \,ii}-selected
  absorbers at $0.07 < z_{\rm abs} < 0.55$.  We detect the 5780\,\AA\ 
  DIB in one line of sight at $z_{\rm abs} = 0.1556$; this is only the
  second QSO absorber in which a DIB has been detected. Unlike the
  majority of local DIB sight-lines, both QSO absorbers with detected
  DIBs show weak 6284\,\AA\ absorption compared with the 5780\,\AA\ 
  band.  This may be indicative of different physical conditions in
  intermediate redshift QSO absorbers compared with local galaxies.
  Assuming that local relations between the 5780\,\AA\ DIB strength
  and \nhi\ and \ebmv\ apply in QSO absorbers, DIB detections and
  limits can be used to derive \nhi\ and \ebmv.  For the one absorber
  in this study with a detected DIB, we derive \ebmv\ = 0.23\,mag and
  $\log$\nhi\ $\ge$ 20.9, consistent with previous conclusions that
  Ca{\sc \,ii} systems have high H{\sc \,i} column densities and
  significant reddening.  For the remaining 8 Ca{\sc \,ii}-selected
  absorbers with 5780\,\AA\ DIB non-detections, we derive \ebmv\ upper
  limits of 0.1--0.3\,mag.
\end{abstract}


\section{Introduction}

Damped Lyman alpha (DLA) systems are usually considered to be the
class of QSO absorber with the highest neutral hydrogen column
densities [\nhi\ $\ge 2 \times 10^{20}$ \cm].  Nonetheless, the DLAs
are characterised by generally low metallicities and gas-phase
depletion fractions
\citep[e.g.][]{KhareP_04a,AkermanC_05a,ProchaskaJ_07a} and low
reddening due to dust (\citealt{MurphyM_04c};
\citealt*{EllisonS_05a}).  The DLAs are also poor in molecules, as
demonstrated by both their generally low fractions of H$_2$
\citep[e.g.][]{LedouxC_03a} and the lack of a detection for any other
molecular species, such as OH or CO \citep[e.g.][]{CurranS_06a}.
Although the handful of DLAs which do exhibit molecular H$_2$
absorption may be biased, e.g.  towards high metallicities
\citep{PetitjeanP_06a}, such systems can offer a novel insight into
the physical conditions of the galactic interstellar medium
\citep[ISM, e.g.][]{SrianandR_05a,NoterdaemeP_07a}.  In addition to
the study of H$_2$, one avenue that is just starting to be explored is
how the diffuse interstellar bands \citep[DIBs; see reviews
by][]{HerbigG_95a,SarreP_06a} may be used to probe the intermediate
redshift ISM.  Although lacking definitive identifications, the
strength (both absolute and relative) of these broad absorption
features in the Milky Way (MW) and other nearby galaxies exhibit
dependencies on (and sometimes, tight correlations with) neutral gas
content, dust reddening, metallicity and local radiation field
\citep[e.g.][]{HerbigG_93a,CoxN_06a,WeltyD_06a,CoxN_07a}.  Moreover,
if DIBs are as strong in DLAs as they are in the MW [i.e.~for a
given \nhi], then they should be relatively easy to detect at
intermediate redshifts.

The first systematic search for DIBs in DLAs has recently been carried
out by Lawton et al.~(in preparation) in 7 $z < 1$ absorbers.  In only
one case were DIBs detected: the 4428, 5705 and 5780\,\AA\
features\footnote{We cite all DIBs with reference to their normal air
  wavelengths, although their vacuum values have been used in practice
  in order to be consistent with our spectral wavelength calibration;
  see Section 2.} were all detected in the $z \sim 0.5$ DLA towards AO
0235$+$164 \citep{JunkkarinenV_04a,YorkB_06a}.  Lawton et al.~showed
that for the 6 non-detections in their DLA sample, the strength of the
5780\,\AA\ DIB [which shows one of the tightest correlations with
\nhi\ in the MW] is often at least 3 times weaker in DLAs for a given
\nhi\ compared with Galactic sight-lines.  The 6284\,\AA\ DIB is even
more under-abundant in DLAs for a given \nhi: 4--10 times weaker than
towards Galactic sight-lines.  A similar result has been found for DIBs
in the Large and Small Magellanic Cloud \citep[LMC and
SMC;][]{WeltyD_06a} where the 5780\,\AA\ DIB is typically 10--30 times
weaker than expected from the Galactic relation.  On the other hand,
the 5780\,\AA\ DIB strength correlates well with \ebmv\ in both
Galactic and Magellanic Cloud sight-lines, and the detection towards AO
0235$+$164 also lies on the same relationship \citep{YorkB_06a}.
These results hint that DIB formation/survival and high dust content
are closely linked and that DIBs are therefore most likely to be
detected in galaxies with high reddening.

\citet*{WildV_06a} have recently suggested that absorbers identified
via high equivalent widths (EWs) of Ca{\sc \,ii} may select the
highest \nhi\ and highest \ebmv\ absorbers.  For example, whereas DLAs
have been constrained to have \ebmv\ $\le$ 0.04
\citep{MurphyM_04c,EllisonS_05a}, \citet{WildV_06a} find that
absorbers with Ca{\sc \,ii} $\lambda$3934 EWs $>$0.7\,\AA\ have
\ebmv\ values up to $\sim$0.1 mag.  Ca{\sc \,ii} absorbers may
therefore be promising sites for the detection of DIBs.

\section{Target Selection, Observations and DIB Search}

\citet{WildV_05a} presented a sample of $0.8 < z < 1.3$ Ca{\sc \,ii}
absorbers selected from the Sloan Digital Sky Survey (SDSS).  However,
the typical rest wavelengths of the strong DIB features (approximately
4500--7000\,\AA) makes the \citet{WildV_05a} sample unsuitable for an
optical search for the diffuse bands.  We have recently conducted an
independent search for Ca{\sc \,ii} absorbers in the SDSS at $z<0.6$
\citep[see][for details]{ZychB_07a} and found over 40 new absorbers.
We selected 9 high-EW (Ca{\sc \,ii} $\lambda$3934 EW$\ge$0.35\,\AA)
systems whose redshift places at least one of the strong DIBs
(specifically, the 4428, 5705, 5780, 5797, 6284 and 6613\,\AA\ bands
were targeted) in regions of the spectrum free from atmospheric
absorption and night sky emission.  We note that due to the small
impact parameters between the QSO and the galaxy causing absorption
\citep[typically $<$ 10 kpc, e.g.][]{ZychB_07a} and the relatively
large contribution by galactic light in the SDSS fibre, it is probable
that some of the Ca{\sc \,ii} EW is contributed by galactic
photospheric absorption.  The SDSS Ca{\sc \,ii} EWs may therefore not
be an accurate measure of the interstellar Ca{\sc \,ii} content (this
effect may be minimised when the Ca{\sc \,ii} EW can be measured from
the FORS long slit spectrum; see Table \ref{dib_table}).  However, the
small impact parameters also mean that these galaxies are likely to
produce significant absorption from their ISM in the QSO spectrum and
the potential contamination from photospheric absorption does not
alter the conclusions of this work [e.g.~inferred \nhi\ and \ebmv].
Table \ref{obs_table} lists the targets\footnote{In Table
  \ref{obs_table} we give the full SDSS identification for each QSO
  but elsewhere we use abbreviated names.} in our sample, as well as
their emission and absorption redshifts, $z_{\rm em}$ and $z_{\rm
  abs}$ (the latter being derived from a Gaussian fit to the Ca~II
$\lambda$ 3934 \AA\ line), and SDSS $r$-band magnitudes, $m_r$.

\begin{table*}
\begin{center}
\begin{minipage}{0.85\textwidth}
\caption{Targets and observational setup.  S/N values are given per
  pixel for the regions in which the DIBs were located.}
\vspace{-2mm}
\begin{tabular}{lcccccccc}
\hline
QSO & $z_{\rm em}$ & $z_{\rm abs}$ & $m_r$  & Grism & Resolution & Coverage & Exp. time & S/N  \\
    &              &               & [mag]  &       & [\AA]      & [\AA]    & [s]       &      \\ \hline
SDSS J001342.44$-$002412.6 & 1.644 & 0.1556 & 18.6 & 1200R & 2.3 & 5960--7370 & 5280 & 60--80  \\
SDSS J100943.55$+$052953.8 & 0.942 & 0.3862 & 16.9 & 600RI & 5.2 & 5495--8620 & 3960 & 130--180 \\
                   & &        &      & 1028z & 2.7 & 7930--9575 & 2640 & 70--130 \\
SDSS J104029.94$+$070528.3 & 1.532 & 0.2063 & 18.6 & 600I & 4.0 & 6925--9470& 2640 & 40--55 \\
SDSS J113702.03$+$013622.1 & 1.641 & 0.4492 & 18.6 & 600RI & 5.2 & 5495--8620 & 5280 & 55--75\\
SDSS J121911.23$-$004345.5 & 2.293 & 0.4485 & 18.0 & 600RI & 5.2 & 5495--8620 & 2640 & 90--115\\
SDSS J122608.02$-$000602.2 & 1.125 & 0.5179 & 18.4 & 600RI & 5.2 & 5495--8620 & 5280 & 100--120\\
SDSS J143701.20$-$010418.0 & 0.286 & 0.0725 & 19.1 & 1200R & 2.4 & 5960--7370 & 10560 & 70--85\\
SDSS J213502.45$+$103823.5 & 1.511 & 0.0984 & 18.8 & 600B & 4.8 & 3490--6360 & 13200 & 70--85\\
SDSS J225913.74$-$084419.6 & 1.290 & 0.5293 & 18.4 & 600RI & 5.2 & 5495--8620 & 5280 & 75--100\\
\hline 
\end{tabular}\label{obs_table}
\end{minipage}
\end{center}
\end{table*}

The 9 targets in Table \ref{obs_table} were observed in long slit mode
with the FORS2 spectrograph on the Very Large Telescope (VLT) in Chile
during ESO's Period 77 (1 April 2006 -- 30 September 2006).
Observations were obtained through a 1 arcsec slit with the CCD binned
2$\times$2.  Grism selection depended on absorber redshift; the
exposure time, choice of grism and the resulting FWHM
resolutions\footnote{The FWHM resolution was calculated as an average
  across the wavelength range based on Gaussian fits to unresolved sky
  lines and 2D arc frames.}  and signal-to-noise (S/N) ratios per
pixel are listed in Table \ref{obs_table}.  The data reduction
procedure followed standard steps for long slit spectra using IRAF: a
median bias frame was subtracted from each science frame, followed by
division by an average lamp flat field. The spectra were optimally
extracted, wavelength calibrated by use of a CuAr lamp and converted
to a vacuum-heliocentric scale.  We experimented with different
methods of combining individual exposures, including the usual
SCOMBINE task in IRAF with weightings according to S/N, and also using
UVES\_popler, as described in \citet{ZychB_07a}.  Both gave very
similar results.

We searched the final spectra for absorption associated with the 4428,
5705, 5780, 5797, 6284 and 6613\,\AA\ diffuse bands.  The first of
these bands is intrinsically broad with an average (rest-frame) FWHM
measured from 4 Galactic stellar sight-lines of FWHM $\sim$12.3\,\AA\ 
\citep{JenniskensP_94a}.  The other 5 DIBS are narrower, with FWHM
values of $\sim$2.2, 2.1, 1.0, 2.6 and 1.1\,\AA\ respectively in Galactic
sight-lines \citep{JenniskensP_94a}\footnote{These values are in good
  agreement with the slightly newer compilation of
  \citet{TuairisgS_00a}, with the exception of the 4428\,\AA\ DIB which
  is reported to have an average (over 3 reddened Galactic lines of
  sight) of FWHM of 17.5\,\AA.}.  Comparing these values with the
resolution of our spectra in Table \ref{obs_table} it can be seen that
the 4428\,\AA\ DIB is always resolved in our spectra.  We usually do
not resolve the narrower DIBs; taking into account the $1+z$
broadening, the expected FWHM values of the DIBs is 2--3\,\AA,
compared with our typical spectral resolution of 3--5\,\AA.

\begin{table*}
\begin{center}
\begin{minipage}{\textwidth}
\caption{DIB and metal line rest-frame EWs and 3$\sigma$ limits. No limit
  is quoted when the line is not covered by the spectrum or is in a
  region of bad sky contamination. Ca{\sc \,ii} values were measured
  from the FORS spectra when possible, otherwise SDSS values are quoted (F
  and S flags respectively). The reddening, \ebmv, is inferred, not
  measured (see text).}
\vspace{-2mm}
\begin{tabular}{lcccccccccc}
\hline
QSO &  $z_{\rm abs}$ & $\lambda$4428 & $\lambda$5705 & $\lambda$5780  & $\lambda$5797 & $\lambda$6284 & $\lambda$6613 & Ca{\sc \,ii} H and K                        & Na{\sc \,i} D$_1$ and D$_2$                       & \ebmv\           \\
    &                & [m\AA]        & [m\AA]        & [m\AA]         & [m\AA]        & [m\AA]        & [m\AA]        & $\lambda$3934, $\lambda$3969 [\AA]   & $\lambda$5891, $\lambda$5897 [\AA]   & [mag]            \\ \hline
J0013$-$0024 & 0.1556 & ...          & $<$62         & 94$\pm$16      & $<$45         & $<$78         & ...           & 1.09$\pm$0.18, 1.34$\pm$0.17 (S) & 1.12$\pm$0.02, 0.94$\pm$0.02 & $\phantom{<}0.23$\\
J1009$+$0529 & 0.3862 & $<$124       & $<$41         & $<$42          & $<$42         & $<$58         & ...           & 0.51$\pm$0.05, 0.19$\pm$0.05 (S) & ...                          &          $<$0.12 \\
J1040$+$0705 & 0.2063 & ...          & ...           & $<$127         & $<$127        & $<$155        & $<$146        & 0.61$\pm$0.12, 0.25$\pm$0.15 (S) & $<0.10$, $<0.10$             &          $<$0.30 \\
J1137$+$0136 & 0.4492 & $<$176       & $<$81         & $<$82          & $<$82         & ...           & ...           & 0.35$\pm$0.04, 0.14$\pm$0.04 (F) & $<0.09$, $<0.09$             &          $<$0.21 \\
J1219$-$0043 & 0.4485 & $<$120       & $<$58         & $<$60          & $<$60         & ...           & ...           & 0.40$\pm$0.02, 0.32$\pm$0.02 (F) & 0.09$\pm$0.02, $<0.06$       &          $<$0.16 \\
J1226$-$0006 & 0.5179 & $<$114       & ...           & ...            & ...           & ...           & ...           & 0.67$\pm$0.02, 0.42$\pm$0.02 (F) & ...                          &          ...     \\
J1437$-$0104 & 0.0725 & ...          & ...           & $<$53          & $<$61         & ...           & $<$51         & 1.07$\pm$0.20, 0.98$\pm$0.30 (S) & 1.31$\pm$0.02, 0.85$\pm$0.02 &          $<$0.15 \\
J2135$+$1038 & 0.0984 & $<$184       & $<$99         & $<$102         & ...           & ...           & ...           & 0.90$\pm$0.19, 0.43$\pm$0.28 (S) & ...                          &          $<$0.25 \\
J2259$-$0844 & 0.5293 & $<$136       & ...           & ...            & ...           & ...           & ...           & 0.38$\pm$0.02, 0.19$\pm$0.03 (F) & ...                          &              ... \\
\hline 
\end{tabular}\label{dib_table}
\end{minipage}
\end{center}
\end{table*}

\begin{figure}
\centerline{\rotatebox{270}{\resizebox{7cm}{!}
{\includegraphics{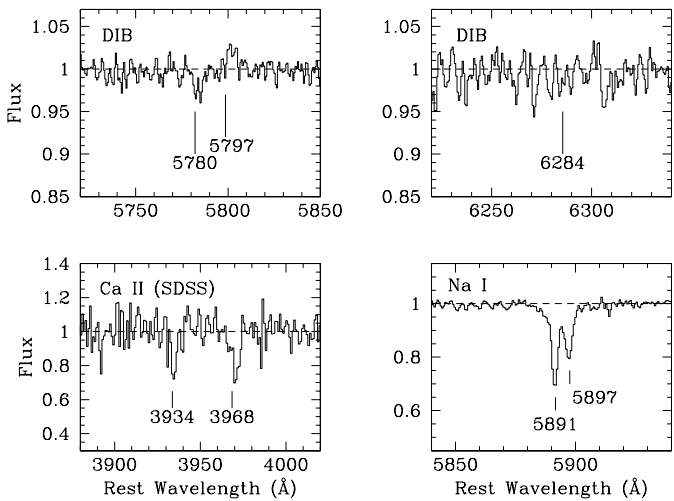}}}}
\caption{\label{dib_fig} Normalised spectra towards J0013$-$0024
covering three of the DIBs of interest in our survey, as well as metal
lines from Ca{\sc \,ii} and Na{\sc \,i}.  All lines are from our FORS spectra,
with the exception of the Ca{\sc \,ii} doublet, where the SDSS spectrum
is shown.  The spectra have
been shifted to a rest wavelength scale for $z=0.1556$.}
\end{figure}

Our search yielded one DIB detection: the 5780\,\AA\ band at $z =
0.1556$ towards J0013$-$0024, the absorber with the highest apparent
Ca{\sc \,ii} EW in our sample. Figure \ref{dib_fig} shows this
detection together with the corresponding Ca{\sc \,ii} (from the SDSS
spectrum) and Na{\sc \,i} absorption.  The redshifts of the 5780 \AA\
DIB and Na{\sc \,i} lines are in excellent agreement with the Ca{\sc \,ii}
absorption.  Depending on the method of weighting the individual
exposures, the velocity offsets between various absorption features 
in the final spectrum are always $<50$ \kms, i.e. less than one 
half of a resolution element; often the agreement is $<$ 20 \kms.  
We measure the EW of the
$\lambda$5780 feature using both a simple integration of optical
depth, as well as by a Gaussian de-blend in order to account for the
presence of a second (unidentified) absorption feature offset by
$\sim$130 \kms\ to the red of the DIB.  The unidentified
feature does not correspond to any known stellar or interstellar features at
$z = 0.1556$ and we conclude that it is likely to be due to gas
at a different redshift.  We also repeat the EW
measurements in the UVES\_popler reduction; all EW values are in
excellent agreement and lie within the statistical 1 $\sigma$ error
derived from the spectral error array.  Our
final quoted EW (see Table \ref{dib_table}) adopts an average of the
EWs determined from various measurement methods and spectral
combinations.

For all other Ca{\sc \,ii} absorbers with non-detections, the
3$\sigma$ detection limits are given in Table \ref{dib_table}.  For
the DIBs that we would expect to be resolved in our spectra (the
4428\,\AA\ band in all cases and, e.g., the 6284\,\AA\ band towards
J1009+0529) we assumed that the absorption would have an observed FWHM
of $1+z$ times the typical Galactic value (see above).  This allowed
us to calculate the number of pixels over which the absorption would
be expected to extend.  For the unresolved lines, we assumed that the
number of pixels was equal to the FWHM spectral resolution (in\,\AA,
see Table \ref{obs_table}) divided by the dispersion (in\,\AA/pixel).

\section{Discussion}

\begin{figure}
\centerline{\rotatebox{0}{\resizebox{9cm}{!}
{\includegraphics{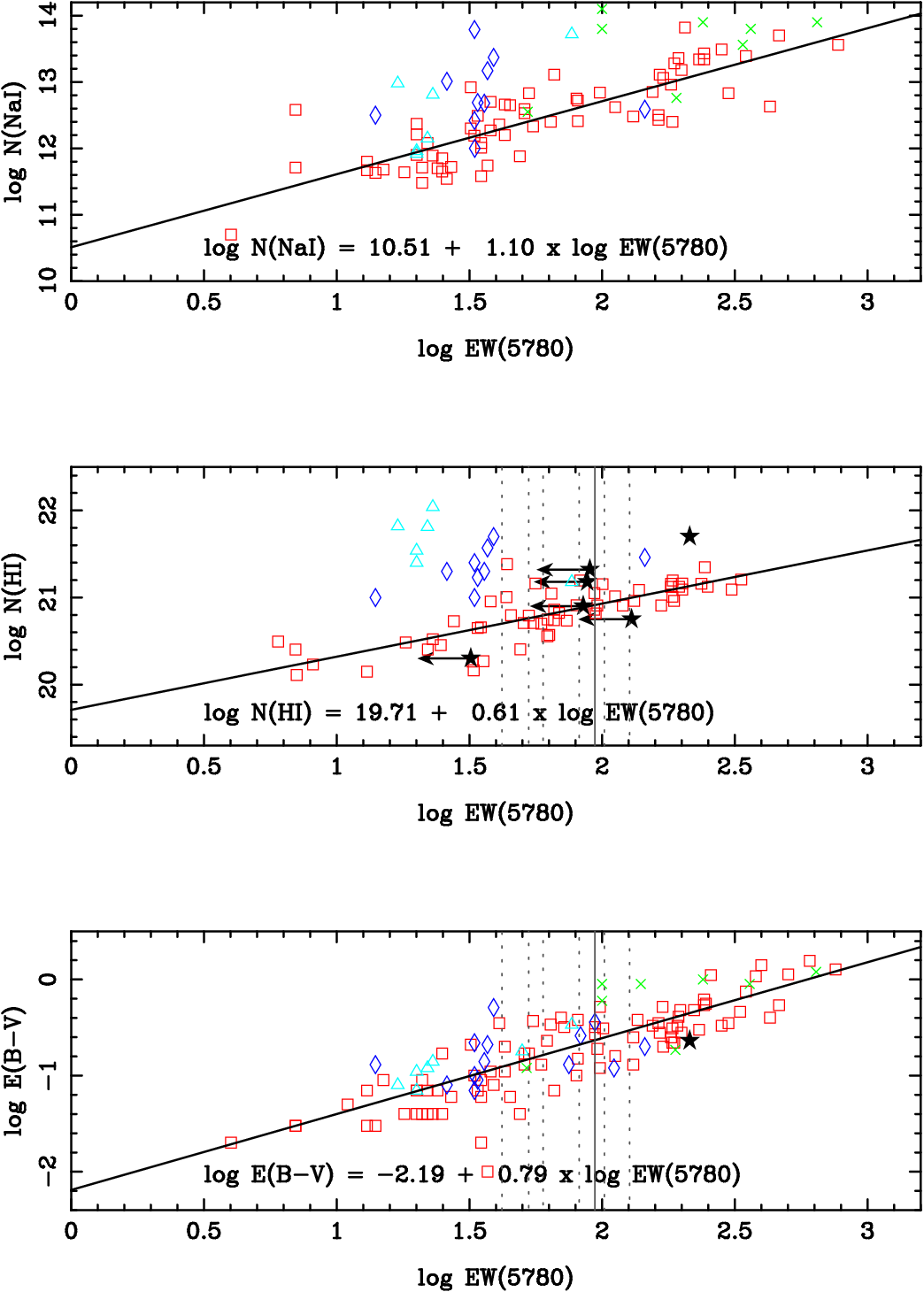}}}}
\caption{\label{ebmv_fig} Correlations of Na{\sc \,i} (top panel),
  \nhi\ (middle panel) and \ebmv\ (bottom panel) versus the log of the
  EW (in m\AA) of the 5780\,\AA\ DIB.  Open squares are Galactic
  points from \citet{HerbigG_93a}; open triangles/diamonds are SMC/LMC
  points from \citet{WeltyD_06a}, \citet{VladiloG_87a} and
  \citet{Vidal-MadjarA_87a}; crosses are other extra-galactic data
  points taken from \citet{SollermanJ_05a}, \citet{DOdoricoS_89a} and
  \citet{HeckmanT_00b}; solid stars are DLAs from \citet{YorkB_06a}
  and Lawton et al.~(in preparation).  The best (least squares) fit to
  the data in each panel is shown with a solid line and the fit
  solution given at the bottom of each panel.  The fit of the
  5780\,\AA\ DIB with \ebmv\ uses all available data; the fits with of
  5780 with \nhi\ and Na{\sc \,i} use just the Galactic data.  In the
  middle and lower panels, the solid vertical line indicates the
  5780\,\AA\ DIB detection towards J0013$-$0024 and the dotted lines
  are the 3$\sigma$ upper limits for 6 other sight-lines where we have
  a 5780\,\AA\ limits (as given in Table \ref{dib_table}).  }
\end{figure}

In local (e.g.~Galactic, LMC, SMC) sight-lines, the 6284\,\AA\
DIB is typically 2--3 times stronger than the 5780\,\AA\ DIB (e.g.,
York et al. 2006a and references therein).  The one exception is the
unusual SMC wing sight-line towards Sk~143 where the 6284\,\AA\ DIB has
an EW less than half that of the 5780\,\AA\ DIB \citep{WeltyD_06a}.
\citet{YorkB_06a} also found that in the one DLA sight-line with a
5780\,\AA\ band detection out of the 7 studied by Lawton et al.~(in
preparation), the EW of the 6284\,\AA\ feature was also constrained to
be less than the EW of the 5780\,\AA\ line. \citet{YorkB_06a} suggested
that these unusual line ratios could be an indication of an ISM that
is more protected from the ambient UV radiation field.  In the Ca{\sc
  \,ii}-selected absorber towards J0013$-$0024, we constrain the EW of
the 6284\,\AA\ DIB to be at least $\sim$20\% weaker than the 5780\,\AA\
band.  The DIB ratios in this absorber are therefore consistent with
those in the DLA detection of \citet{YorkB_06a} and the SMC wing
sight-line Sk~143 but inconsistent with other local sight-lines,
including starburst galaxies \citep{HeckmanT_00b} and the Magellanic
Clouds \citep{WeltyD_06a}.

In the Galaxy, many DIBs show correlations with \nhi\ and $N$(Na{\sc \,i})
\citep[e.g.][]{HerbigG_93a,HerbigG_95a}.  In Table \ref{dib_table} we
tabulate the EWs of the Na{\sc \,i} doublet for our Ca{\sc
  \,ii}-selected absorbers.  However, we do not calculate the Na{\sc
  \,i} column density because, if the lines are strong enough to be
detected in our moderate resolution spectra, they are likely to be
saturated.  The Galactic correlations of \nhi\ and $N$(Na{\sc \,i}) with
the 5780\,\AA\ DIB, which is one of the tightest of the DIB relations,
is shown in Fig. \ref{ebmv_fig}.  We also show data for the
Magellanic Clouds \citep{WeltyD_06a} and DLAs (\citealt{YorkB_06a};
Lawton et al.~in preparation), where it can be seen that the DIBs are
weak for their \nhi\ compared with the Galactic correlation.  As shown
in Fig. \ref{ebmv_fig}, the DIBs in extra-galactic sight-lines are
also weak for their Na{\sc \,i} column densities.  These departures
from the Galactic relations are probably due to a combination of
effects including ambient radiation field, metallicity and dust-to-gas
ratios \citep{CoxN_06a}.  Assuming that the Galactic 5780--\nhi\ 
relation provides a lower limit for the H{\sc \,i} column density, DIB
detections may be useful for constraining \nhi\ in the absence of
\lya\ observations.  For example, \citet{WildV_05a} and
\citet{WildV_06a} have argued that Ca{\sc \,ii} absorbers represent
the high column density end of the DLA distribution.  Our detection of
the 5780\,\AA\ DIB in the $z_{\rm abs} = 0.1556$ absorber towards
J0013$-$0024 supports this hypothesis, and we derive $\log$\nhi\ $\ge$
20.9 for this absorber.

Unlike correlations with \nhi\ and $N$(Na{\sc \,i}),
\citet{WeltyD_06a} have shown that the 5780\,\AA\ DIB strength follows
a single relationship with \ebmv\ in both Galactic and Magellanic
Cloud sight-lines.  \citet{YorkB_06a} found that the single DLA
5780\,\AA\ DIB detection towards AO 0235$+$164 fell on the same
relationship.  It is not yet clear whether the apparent universality
of this correlation is driven by a tight physical connection between
dust properties and DIB formation \citep{CoxN_07a} or whether it is
coincidence of different physical drivers working in different
directions \citep{CoxN_06a}.  However, if the 5780--\ebmv\ is
applicable to QSO absorbers, we can use our DIB detection limits to
constrain their reddening.  \citet{WeltyD_06a} derive a best fit
correlation between the 5780\,\AA\ DIB (in m\AA) and the \ebmv\ for
Galactic sight-lines: log \ebmv\ = $-$2.70 + 1.01 log EW(5780).  We
derive the best fit relation to the 5780--\ebmv\ data points of the
Galactic plus Magellanic Cloud plus AO 0235$+$164 DLA sight-lines and
find log \ebmv\ = $-$2.19 + 0.79 log EW(5780) (see Figure
\ref{ebmv_fig}).  The range in log \ebmv\ values around the best fit
relation is $\sim \pm$ 0.4 dex.  This correlation gives a reddening
for the Ca{\sc \,ii} absorber towards J0013$-$0024 of
\ebmv$\,\,\sim0.23$\,mag and upper limits for the other 8 Ca{\sc \,ii}
absorbers in our sample of 0.1--0.3\,mag.  These values provide
independent estimates of reddening associated with Ca{\sc
  \,ii}-selected absorbers that do not depend directly on the choice
of extinction law and can be applied for individual absorbers and not
just in a statistical fashion
\citep[e.g.][]{MurphyM_04c,WildV_05a,WildV_06a}.  The Ca{\sc \,ii} EWs
of our sample are typically $<0.7$\,\AA\ (see Table \ref{dib_table});
for this range of EWs, \citet{WildV_06a} determine average reddenings
of \ebmv\ = 0.02, 0.03 and 0.03\,mag for MW, LMC and SMC extinction
curves respectively.

\section{Summary and Future Prospects}

We have reported the results from the first search for DIBs towards 9
Ca{\sc \,ii}-selected absorbers in the redshift range $0.07 \le z_{\rm abs} \le
0.55$.  In one case, the $z_{\rm abs} = 0.1556$ absorber towards
J0013$-$0024, we detect the 5780\,\AA\ DIB.  This absorber has the
highest Ca{\sc \,ii} $\lambda$ 3934 EW in our sample, although there
is some contribution from galactic photospheric absorption in the SDSS
EW measurement.  J1437$-$0104 has only a marginally lower Ca{\sc \,ii}
EW, but a 5780\,\AA\ DIB upper limit that is half that of
J0013$-$0024.  J0013$-$0024 is only the second QSO absorber in which
DIBs have been detected.  Assuming that the Galactic relation between
\nhi\ and $\lambda$5780 EW can be used to derive a lower limit for
H{\sc \,i} column density, we find $\log$\nhi\ $\ge$ 20.9.  Similarly,
the correlation between $\lambda$5780 EW and \ebmv\ seen in all
sight-lines (Galactic and extra-galactic) to date implies a high
reddening in this absorber of \ebmv\ = 0.23\,mag.  These results
provide independent support for the suggestion by \citet{WildV_06a}
that the Ca{\sc \,ii} absorbers are amongst the highest \nhi\ and most
highly reddened of the QSO absorbers.  Indeed, the \ebmv\ derived for
J0013$-$0024 is even higher than the typical statistical values
derived by \citet{WildV_06a}.  For the other absorbers in our sample
we derive upper limits to the reddening of $\sim$0.1--0.3\,mag. We
also find the interesting result that, in contrast to essentially
every local sight-line (with the exception of one SMC wing cloud), both
QSO absorbers with detected DIBs have stronger 5780\,\AA\ features
than 6284\,\AA\ features, possibly due to less intense radiation
fields.

What are the prospects for further detections of DIBs in Ca{\sc \,ii}
and other QSO absorption line systems?  If the relationship of the
5780\,\AA\ DIB strength with \ebmv\ is widely applicable, then
targetting the most reddened systems is likely the most profitable
path.  Nonetheless, this is a challenging prospect; even for an
absorber with \ebmv\ = 0.1\,mag, the rest frame $\lambda$5780 EW will
be $\sim$30 m\AA, which would require an improvement in our current
detection limits by typically a factor of 2--3.  The best candidates
for future observations are likely to be the highest EW Ca{\sc \,ii}
and Mg{\sc \,ii} absorbers, since these appear to be the most highly
reddened of the QSO menagerie \citep{WildV_06a,YorkD_06a,MenardB_07a}.
However, with more detections in hand, it would be possible not only
to infer the gas and dust properties of the absorbers, as described
above, but also to obtain new insight into the nature of the DIBs
themselves.  For example, \citet{YorkB_06a} showed that, in the
$z_{\rm abs} \sim 0.5$ DLA towards AO 0235$+$164, the ratio of the
5705/5780\,\AA\ DIBs was consistent with those in Galactic sight-lines
\citep[e.g.][]{ThorburnJ_03a}.  The identification of such DIB
`families' may indicate which bands are associated with the same
molecular carrier and hence provide guidelines for their chemical
identification.  If the Galactic 5705/5780 relation holds for the
Ca{\sc \,ii} absorber towards J0013$-$0024, we would expect to detect
it with an increase of a factor of three in S/N, possible with 
approximately one more night of integration.

\section*{Acknowledgments} 

SLE was funded by an NSERC Discovery grant.  BAY was partially funded
by an NSERC PGS-M award. MTM thanks the STFC for an Advanced
Fellowship, held at the Institute of Astronomy. BJZ and AMS thank
STFC and EPSRC, respectively, for studentships.

%\bibliography{references} \bibliographystyle{mn2e}

\bspsmall

\label{lastpage}

\end{document}